\documentclass[a4paper]{jpconf}
\usepackage{graphicx}
\usepackage{subfig}
\usepackage{array}
\usepackage{multirow}
\usepackage{hyperref}
\begin{document}

\title{Probing highly collimated photon-jets with deep learning}

\author{Xiaocong Ai$^1$, Shih-Chieh Hsu$^2$, Ke Li$^2$, Chih-Ting Lu$^3$}

\address{$^1$ Deutsches Elektronen-Synchrotron DESY, Notkestr. 85, 22607 Hamburg, Germany}
\address{$^2$ Department of Physics, University of Washington}
\address{$^3$ School of Physics, Korea Institute for Advanced Study, Seoul 02455, Republic of Korea}

\ead{xiaocong.ai@desy.de}

\begin{abstract}
Many extensions of the standard model (SM) predict the existence of axion-like particles
and/or dark Higgs in the sub-GeV scale. Two new sub-GeV particles, a scalar and a pseudoscalar, produced through the Higgs boson exotic decays, are investigated. The decay signatures of these two new particles with highly collimated photons in the final states are discriminated from the ones of SM backgrounds using the Convolutional Neural Networks and Boosted Decision Trees techniques. The sensitivities of searching for such new physics signatures at the Large Hadron Collider are obtained.
\end{abstract}

\section{Introduction}
Exploring anomalous particles from beyond standard model (BSM) signatures is one important goal of the Large Hadron Collider (LHC) physics program. Recently, new particles in the sub-GeV scale have received more and more attention. The light pseudoscalar such as axion-like particles (ALPs)~\cite{Bauer:2018uxu} and light scalar such as dark Higgs~\cite{Winkler:2018qyg} are proposed by many BSM models as mediators in sub-GeV dark matter models~\cite{Matsumoto2019,Bondarenko2020}. In addition to fixed target and B factories experiments, the light pseudoscalar and scalar can also be explored at the LHC via the Higgs boson exotic decays and ALP/dark Higgs strahlung processes.

In this study, the Higgs portal model is used as a prototype to study a novel signature with highly collimated photons forming a jet-like structure as photon-jet~\cite{Dobrescu:2000jt,Chang:2015sdy,Domingo:2016yih}. Two new particles, a scalar ($s$) and a pseudoscalar ($a$), which are much lighter than the SM-like Higgs boson ($h$), are introduced. In this work, the case that the light $s$ $(a)$ dominantly decays to $\gamma\gamma$ and/or $2\pi^0$ ($\gamma\gamma$ and/or $3\pi^0$) where $\gamma$ ($\pi^0$) is the photon (neutral pion) is studied. Both $s$ and $a$ can be pair produced from the Higgs boson exotic decays. They are highly boosted after the production of on-shell $h$ such that the final states are two photon-jets with different substructures depending on the decay modes of $s$ and $a$. The relevant signal processes are (i) $gg\rightarrow h\rightarrow ss\,(aa)\rightarrow (\gamma\gamma)(\gamma\gamma)$, (ii) $gg\rightarrow h\rightarrow ss\rightarrow (\pi^0\pi^0)(\pi^0\pi^0)$, (iii) $gg\rightarrow h\rightarrow aa\rightarrow (\pi^0\pi^0\pi^0)(\pi^0\pi^0\pi^0)$.

The signatures of the photon-jets from $s$/$a$ are identified using an ATLAS-like electromagnetic calorimeter~\cite{Aaboud2019} with the electromagnetic showers simulated using GEANT4~\cite{Agnostelli2003}. In particular, the photon-jet signatures with the presence of SM backgrounds such as the single photon or neutral $\pi^0$ originating from QCD jets is studied using the deep learning technique, the Convolutional Neural Networks (CNN)~\cite{Ayyar2020}, and the Boosted Decision Trees (BDT)~\cite{ROE2005577}. The BDT uses specialized shower shape variables to capture the difference of electromagnetic shower development in the electromagnetic calorimeter (ECAL) between signal and backgrounds. Its performance is compared to CNN which constructs high level feature variables starting from the deposited energy per cell of the ECAL therefore can extract the maximum amount of information from the raw measurements of the ECAL. Based on the photon-jet identification performance, the physics sensitivities of searching for the Higgs exotic decays in $pp$ collisions are derived.

\section{Photon-jet Identification with Deep Learning}

\subsection{Simulation}
A lead/liquid-argon (LAr) sampling ECAL with granularity similar to that of the ATLAS electromagnetic calorimeter has been simulated using GEANT4 with a pseudorapidity ($\eta$) coverage $-0.2 < \eta < 0.2$.
It consists of a thin pre-sampling layer and three sampling layers longitudinal in the shower depth.
The first sampling layer is segmented into high-granularity strips in the $\eta$ direction, with a cell size of 0.0031 $\times$ 0.098 in $\Delta \eta \times \Delta \phi$. The pre-sampling layer, second sampling layer and third sampling layer have granularity of 0.025 $\times$ 0.01, 0.025 $\times$ 0.0245 and 0.05 $\times$ 0.0245 in $\Delta \eta \times \Delta \phi$, respectively. 

The interaction of the photon-jet produced through the signal processes, the single photon and $\pi^0\rightarrow \gamma \gamma$ backgrounds with the ECAL and their deposited energies in each cell of the calorimeter are separately simulated with GEANT4. Four benchmark masses of $s$/$a$, i.e.~0.45 GeV, 0.6 GeV, 0.8 GeV and 1 GeV are studied. For each process being studied, sample of 100,000 events with energy of the $s$/$a$ signal or $\gamma$/$\pi^0$ background uniformly distributed from 40 to 250 GeV is generated. The $\eta$ of $s/a$, $\gamma$ and $\pi^0$ is fixed to zero. Out of each sample, 70\% of the events is used as the training set and the rest is used as the test set for performance evaluation.

\subsection{Separation of photon-jet signatures from SM backgrounds using CNN}
\begin{figure}
\begin{center}
\subfloat[]{\includegraphics[width=.33\columnwidth]{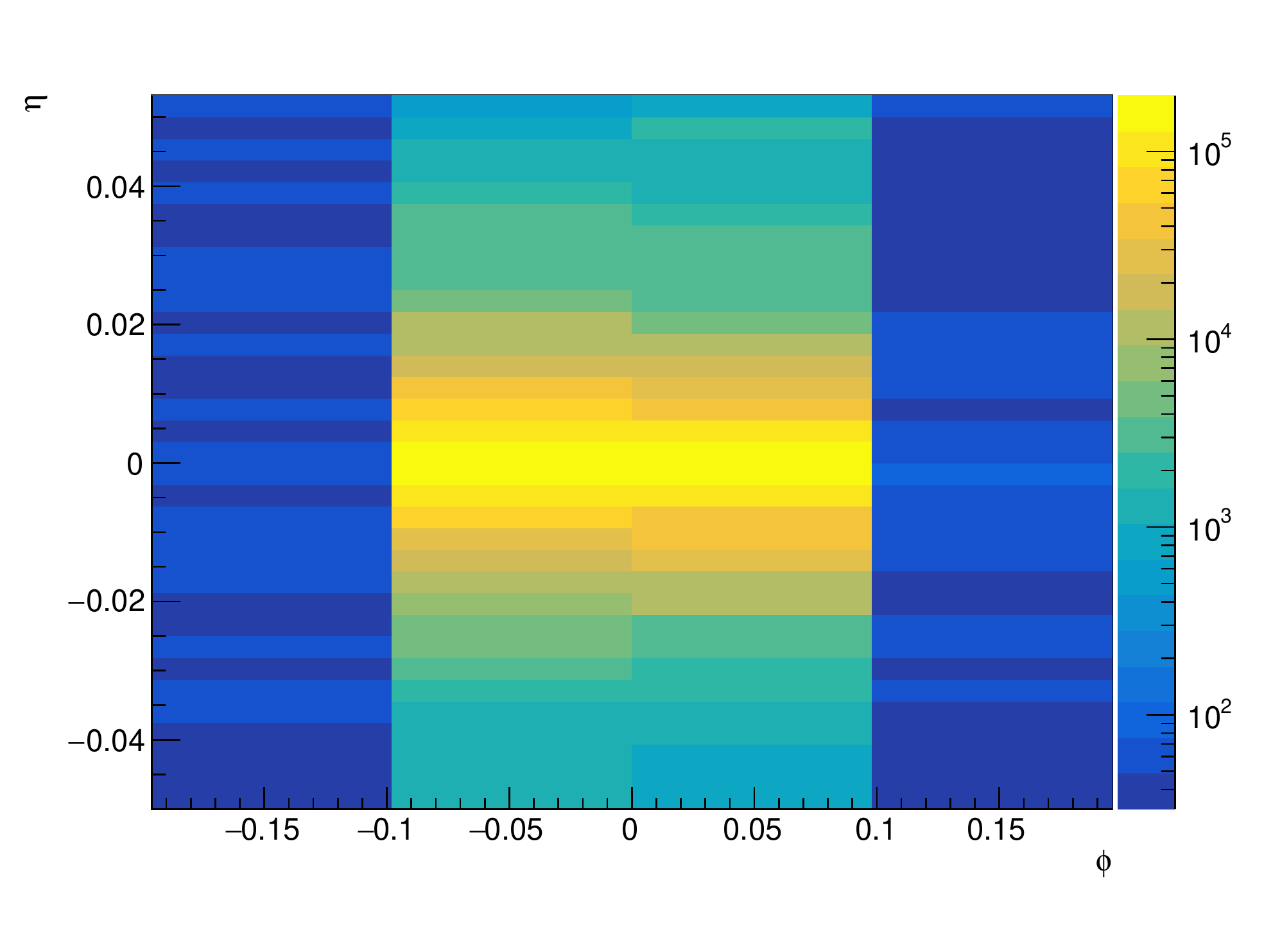}}
\subfloat[]{\includegraphics[width=.33\columnwidth]{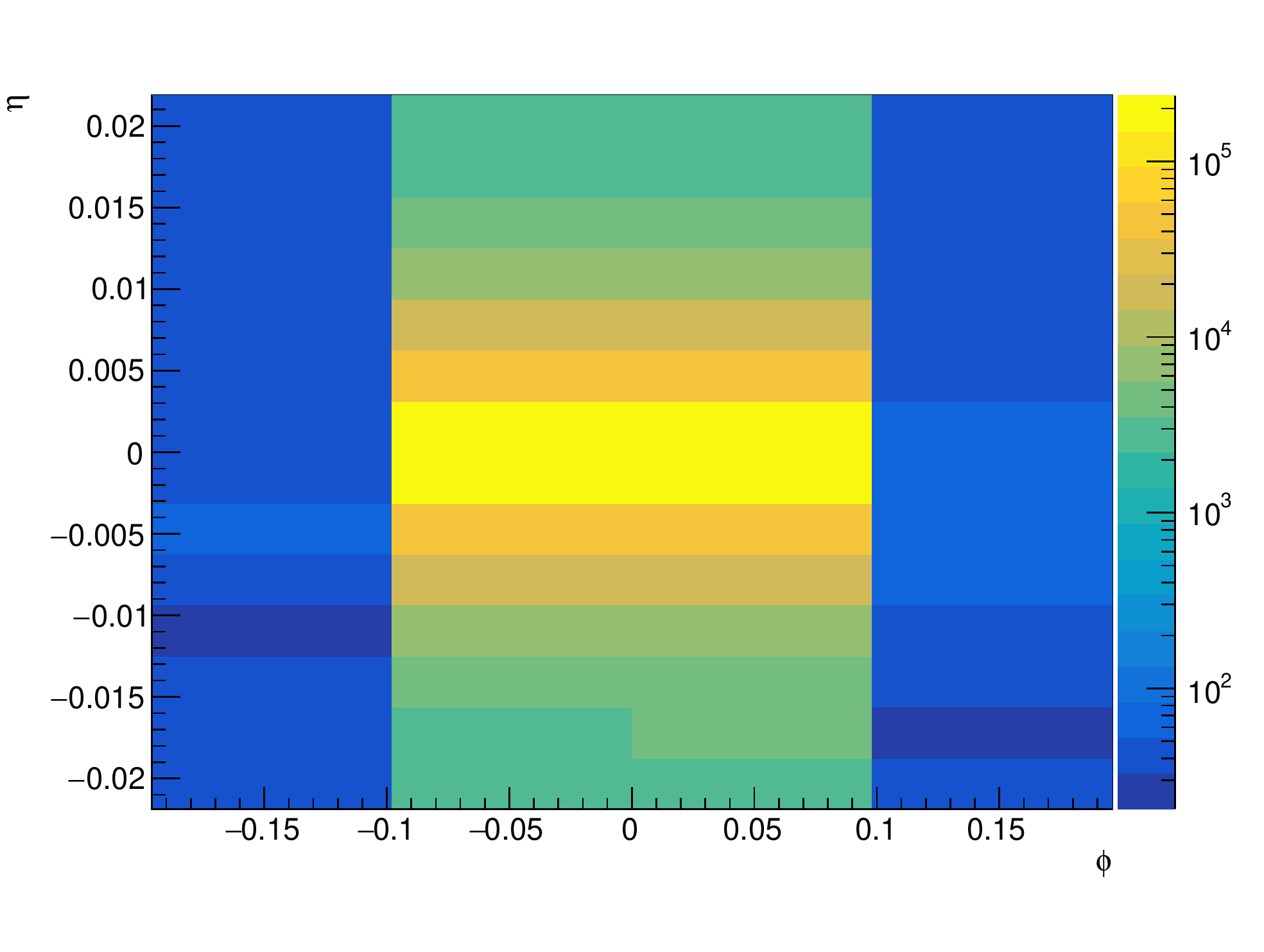}}
\subfloat[]{\includegraphics[width=.33\columnwidth]{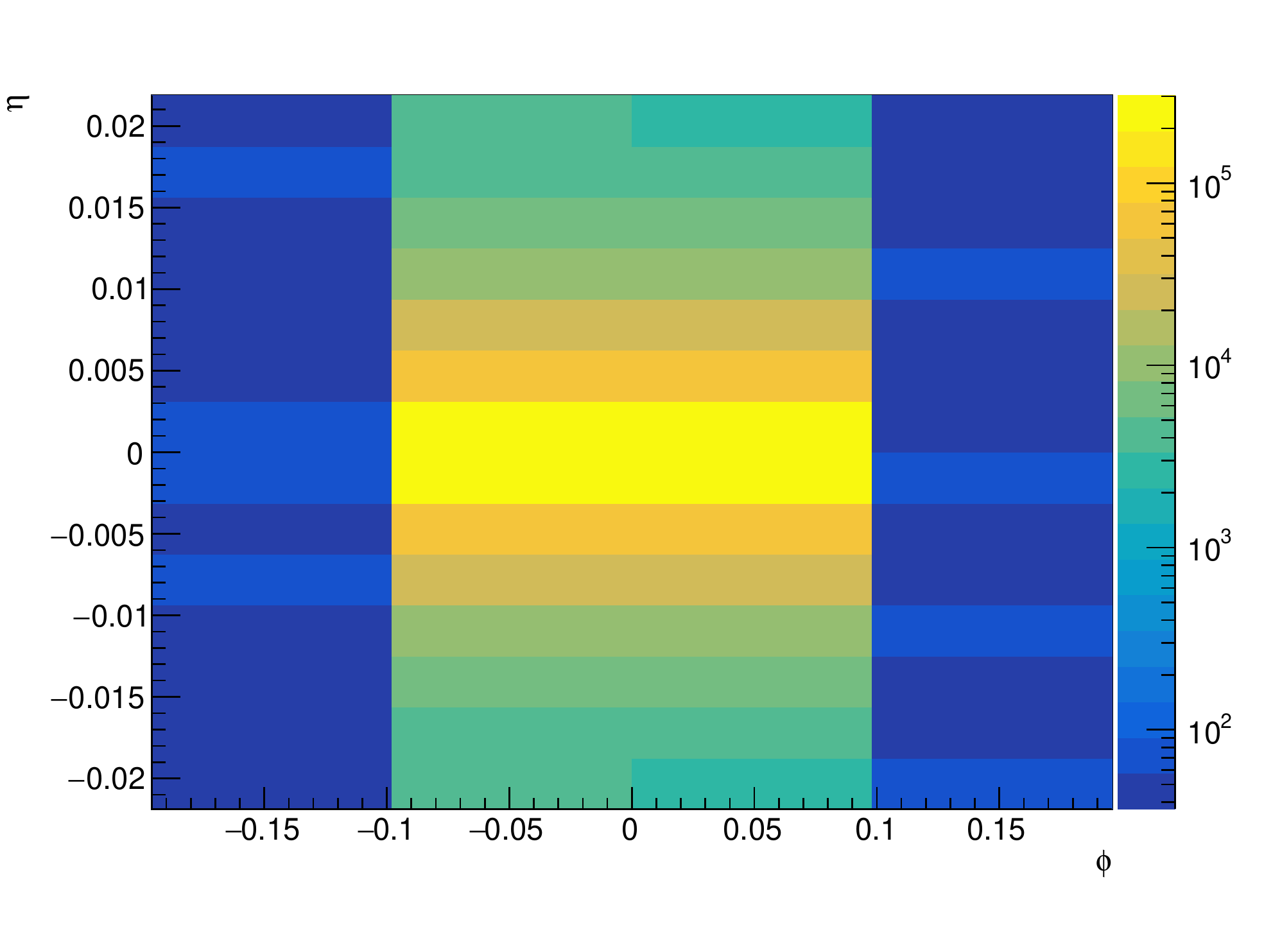}}
\end{center}
\caption{\label{fig:ene_deposits}The deposited energy per cell of (a) photon-jet from $a\rightarrow \gamma \gamma$ ($m_a = 1$ GeV) (b) $\gamma$ (c) $\pi^0$ at the first layer of the ECAL. The $a$, $\gamma$ and $\pi^0$ have energy in the range of [40, 250] GeV.}
\end{figure}

As shown in Fig.~\ref{fig:ene_deposits}, the energy deposits from all cells per ECAL layer are represented as a 2D image of dimensions $N_{cells} (\eta) \times N_{cells} (\phi)$, where $N_{cells} (\eta)$ and $N_{cells} (\phi)$ are the number of cells in $\eta$ and $\phi$ direction, respectively. The value for each cell represents the energy deposited in it.

The {\it Keras}~\cite{chollet2015keras} package with {\it Tensorflow}~\cite{tensorflow2015-whitepaper} as backend is used for implementing the CNN.
Four separated CNN models have been built for the four ECAL layers. Each CNN model is constructed with two convolutional layers with filters of size $3\times3$ and stride 1 and the \textit{rectified linear unit} (RELU)~\cite{agarap2019deep} activation function. Each convolutional layer is followed by a maxpooling layer of size $2\times2$. A flatten layer is used to convert the 2D output array from the pooling layer to 1D array. The 1D arrays from the four CNN models are concatenated before fed to two fully connected layers with 32 nodes and the RELU activation. The final output layer is a fully connected layer with the \textit{softmax} activation function and number of nodes equal to the number of distinct classes providing a multi-class output. 
The categorical \textit{cross-entropy} loss function with the \textit{adam} optimizer~\cite{kingma2017adam} is applied.

\begin{figure}
\begin{center}
\subfloat[]{\includegraphics[width=.3\columnwidth]{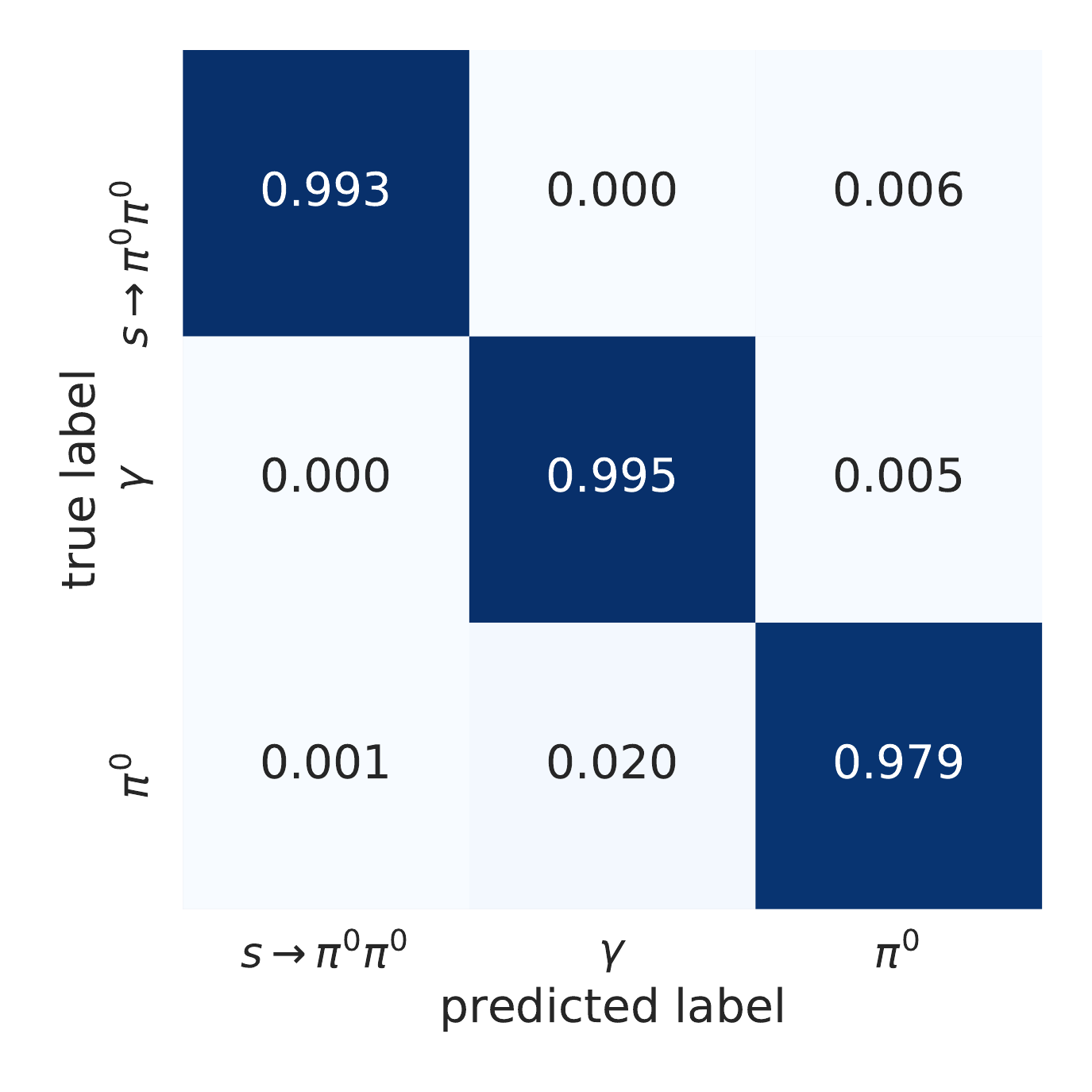}}
\subfloat[]{\includegraphics[width=.3\columnwidth]{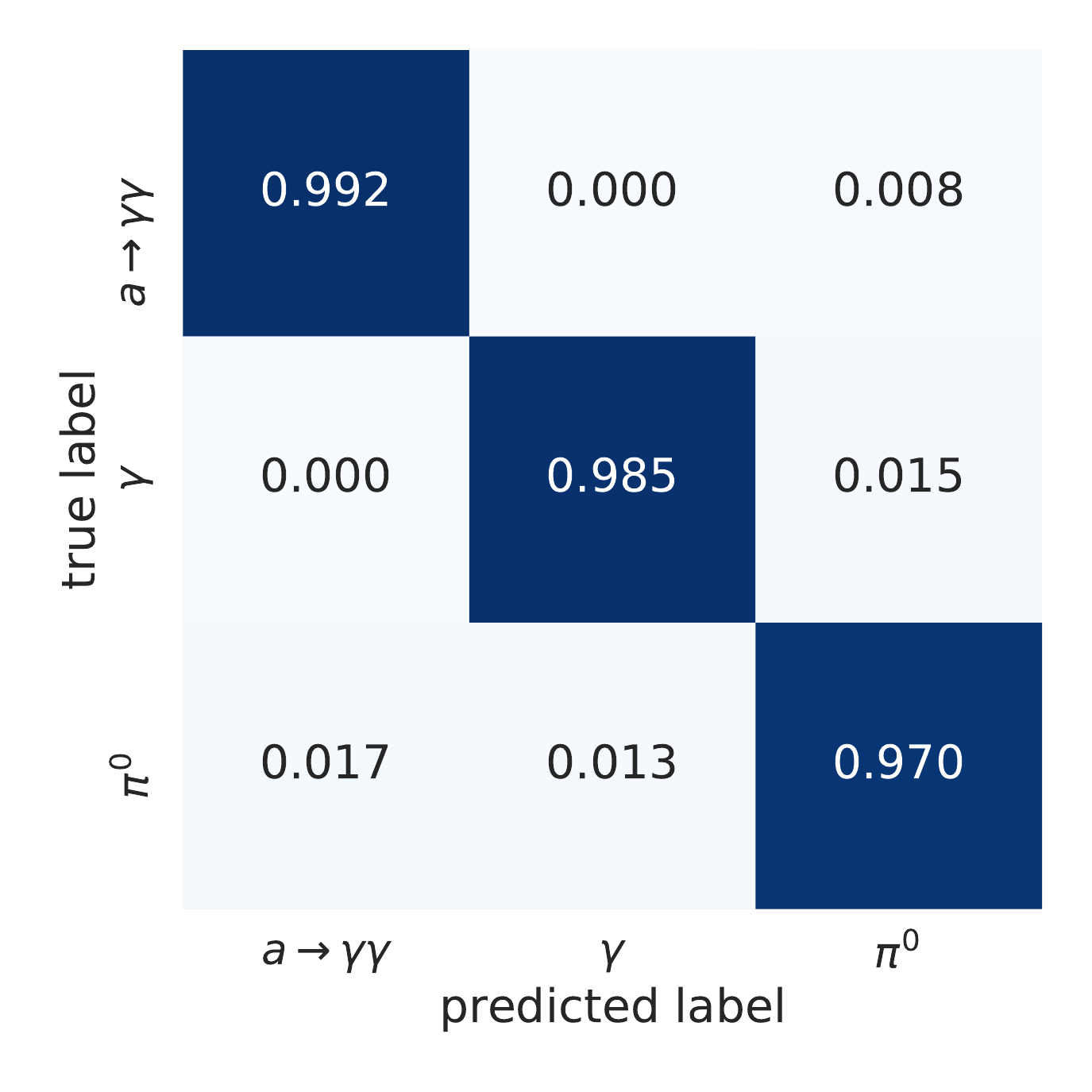}}
\subfloat[]{\includegraphics[width=.3\columnwidth]{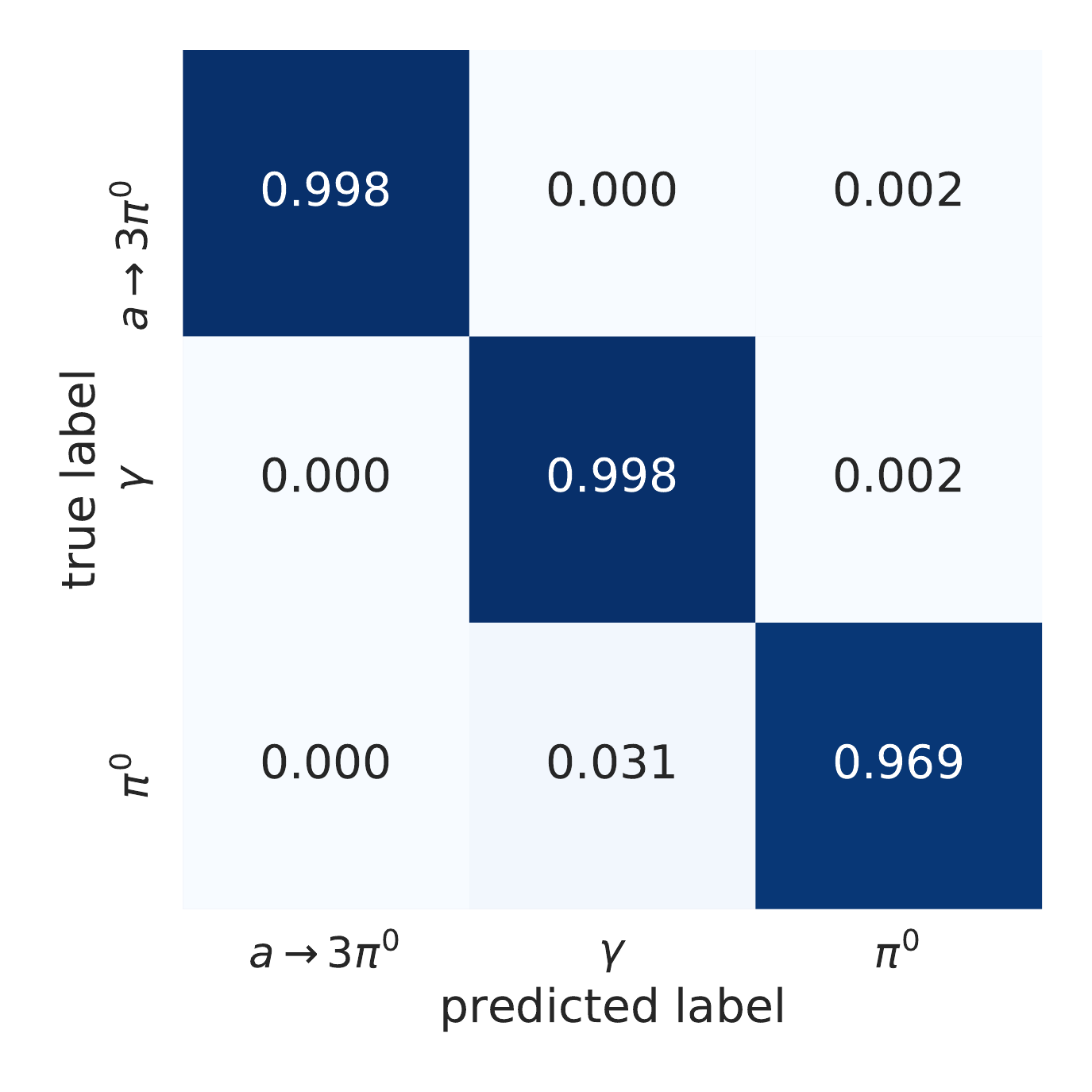}}
\end{center}
\caption{\label{fig:CNN_confmatrix}The normalized confusion matrix for distinguishing between the photon-jet produced in the process of (a) $s\rightarrow \pi^{0} \pi^{0}$ (b) $a\rightarrow \gamma \gamma$ (c) $a\rightarrow \pi^{0} \pi^{0} \pi^{0}$, and the single photon and $\pi^0$ backgrounds for the test set using CNN. The mass of $s$ and $a$ is assumed to be 1 GeV.}
\end{figure}

The confusion matrix for distinguishing between the photon-jet produced in three different signal processes and the same $\gamma$, $\pi^0$ backgrounds with the test set using CNN are shown in Fig.~\ref{fig:CNN_confmatrix}. The overall identification efficiency of the photon-jet is above 99.2\% with $\gamma$ ($\pi^0$) rejection rate above 99.9\% (99.8\%).

\subsection{Comparison of performance between CNN and BDT}
The Gradient BDT~\cite{friedman2001greedy} is used for binary classification of the signal and backgrounds based on the shower shape variables as used in Ref.~\cite{Aaboud2019}. The BDT is trained in four bins of the energy, [40, 100] GeV, [100, 150] GeV, [150, 200] GeV and [200, 250] GeV.

\begin{figure}
\begin{center}
\subfloat[]{\includegraphics[width=.33\columnwidth]{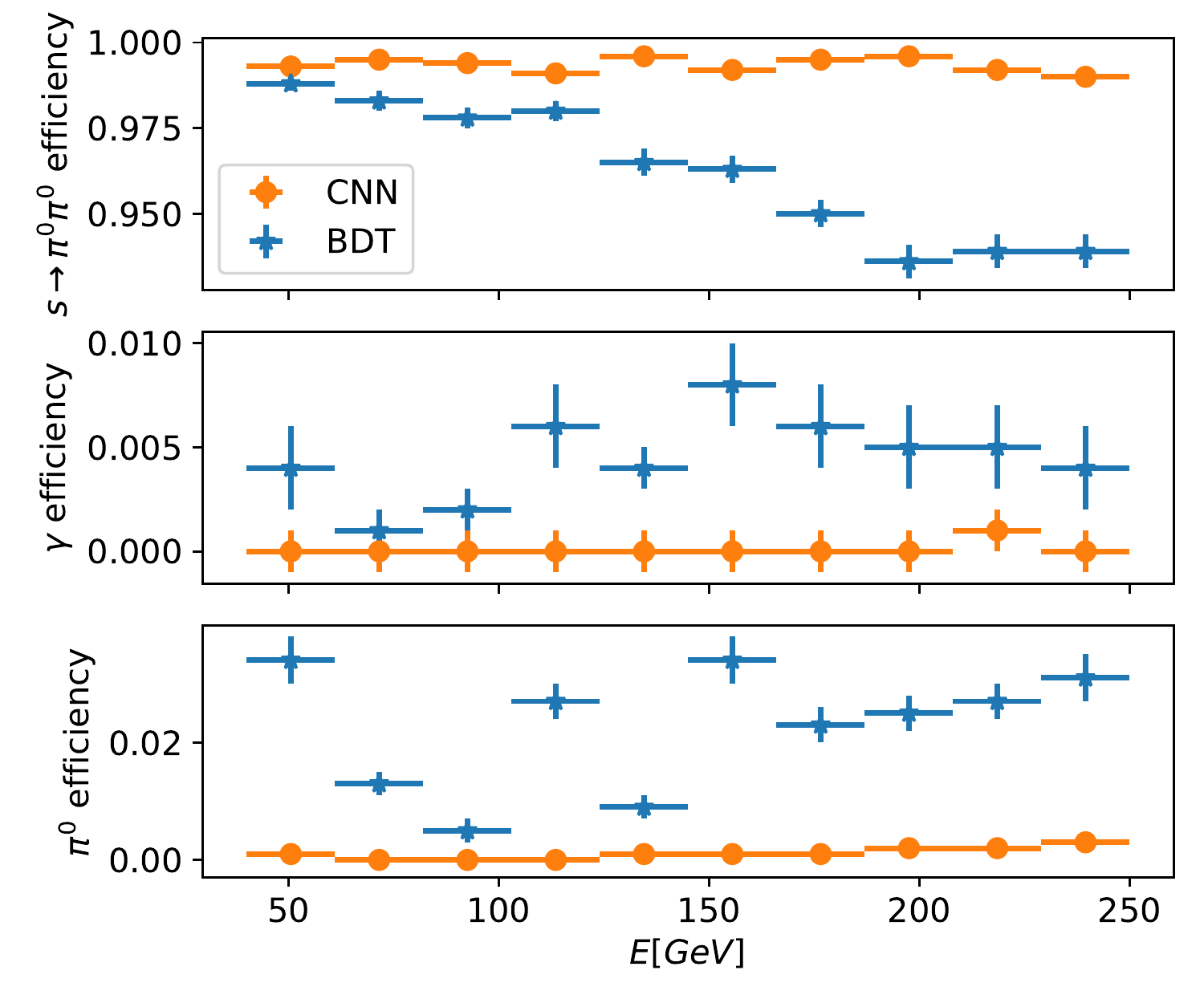}}
\subfloat[]{\includegraphics[width=.33\columnwidth]{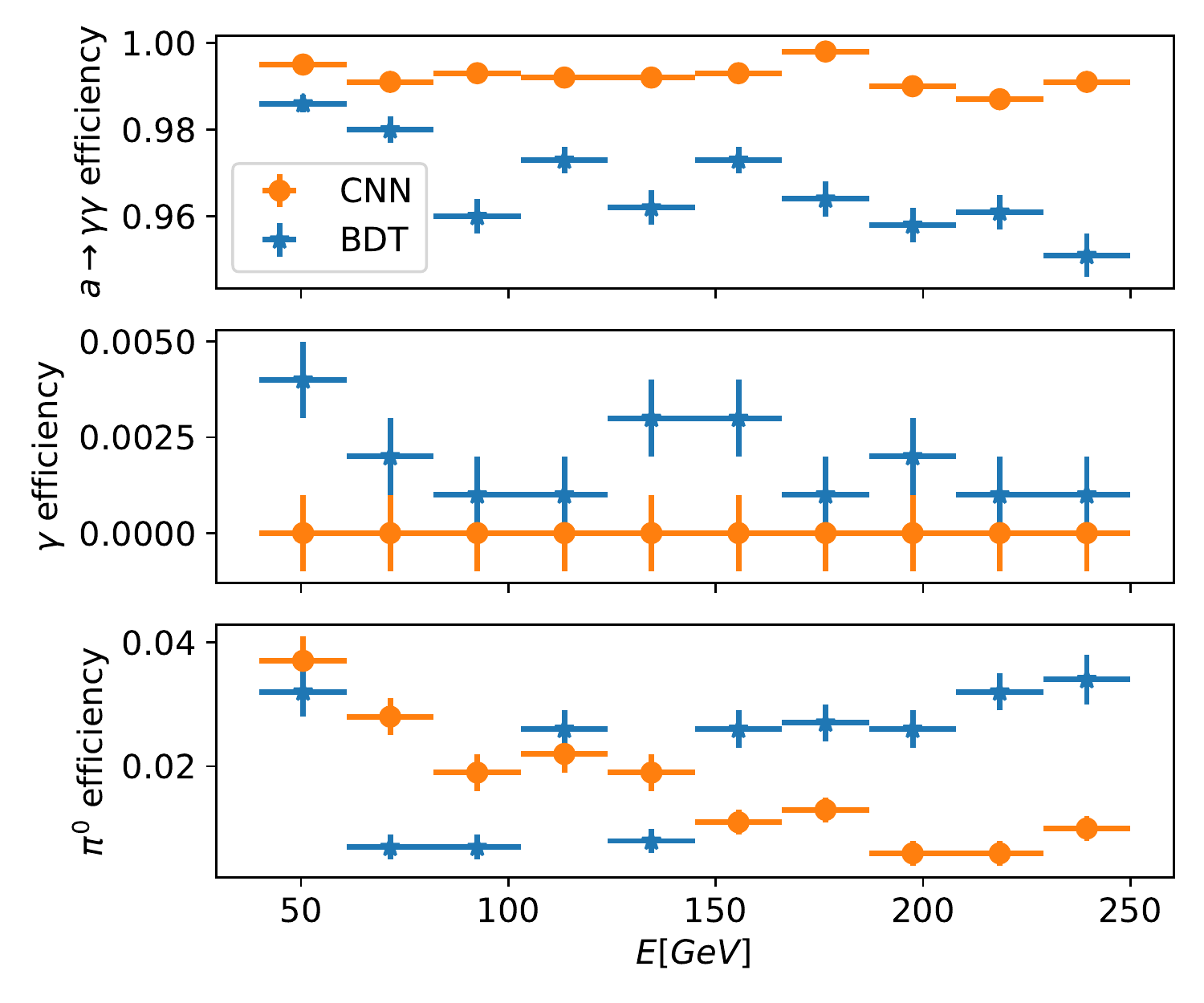}}
\subfloat[]{\includegraphics[width=.33\columnwidth]{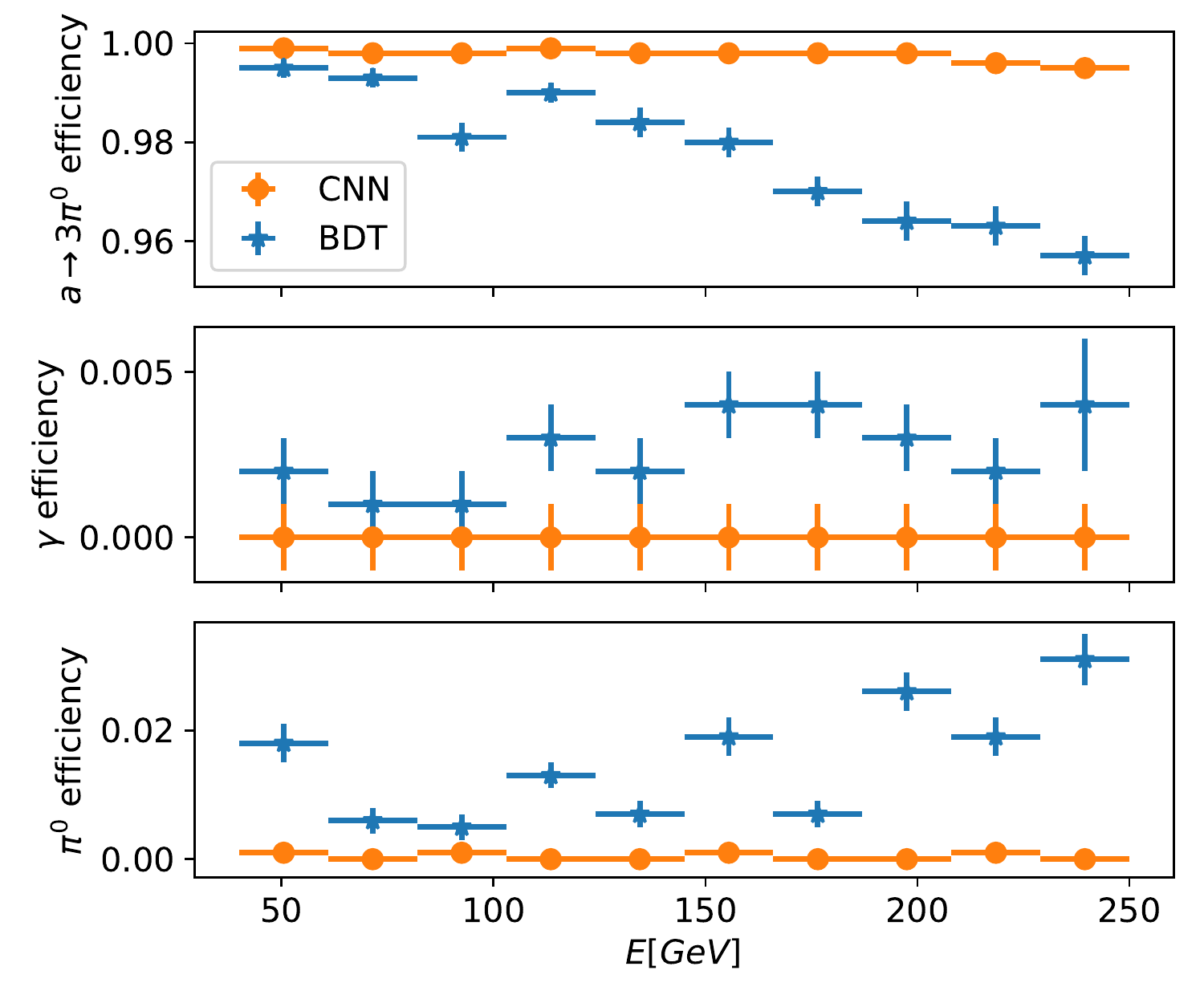}}
\end{center}
\caption{\label{fig:comp_eff_vs_ene} The comparison of identification efficiency of photon-jet (top panels) produced in the process of (a) $s\rightarrow \pi^{0} \pi^{0}$ (b) $a\rightarrow \gamma \gamma$ (c) $a\rightarrow \pi^{0} \pi^{0} \pi^{0}$ and the single photon (middle panels) and $\pi^0$ (bottom panels) backgrounds as a function of energy between CNN (orange dots) and BDT (blue triangles) using the test set. The mass of $s$ and $a$ is assumed to be 1 GeV.}
\end{figure}

Figure~\ref{fig:comp_eff_vs_ene} shows the comparison of the identification efficiency of the photon-jet signal and $\gamma$/$\pi^0$ background as a function of their energy between CNN and BDT. Compared to BDT, CNN has higher photon-jet identification efficiency, in particular at energy above 150 GeV, and higher $\gamma$ rejection rate. When photon-jet is produced through $a\rightarrow \gamma \gamma$, CNN has slightly lower $\pi^0$ rejection rate than BDT.

\section{Physics Sensitivity}

\begin{figure}
\begin{center}
\subfloat[]{\includegraphics[width=.4\columnwidth]{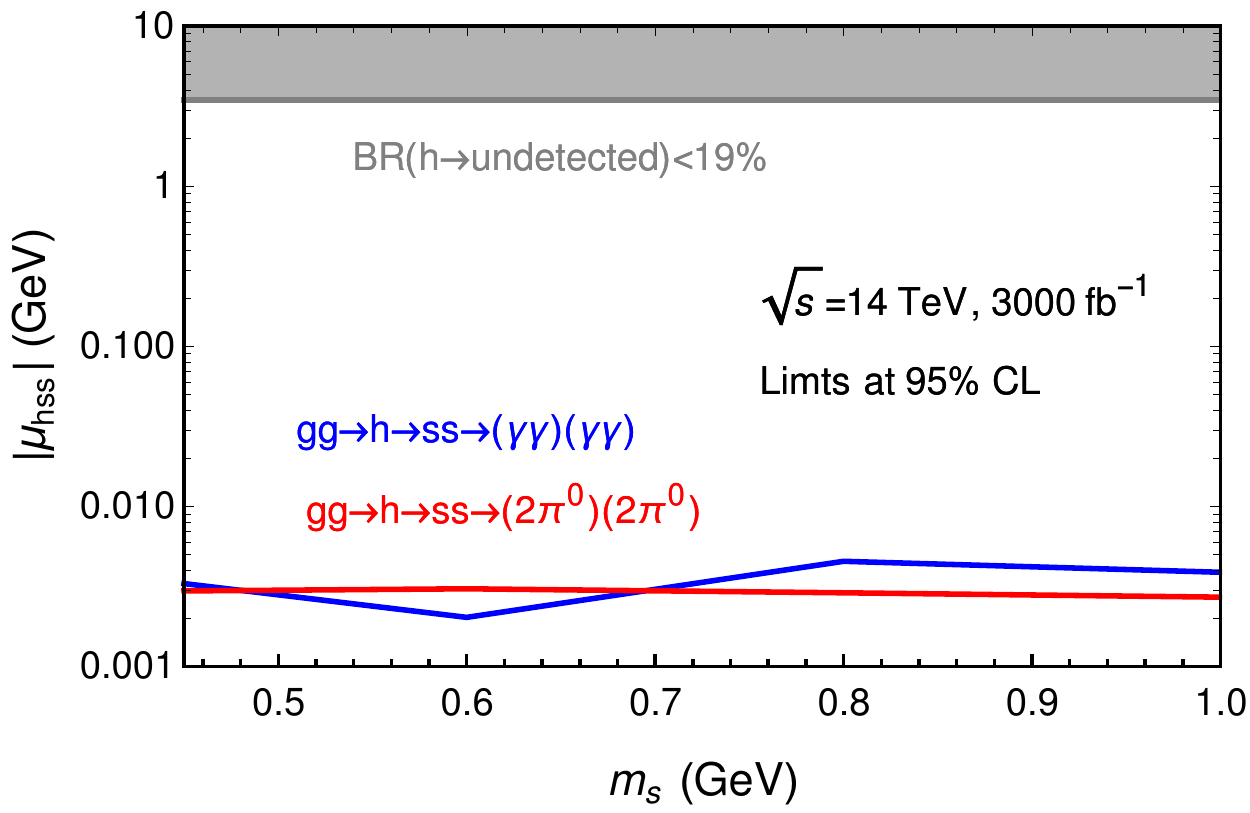}}
\subfloat[]{\includegraphics[width=.4\columnwidth]{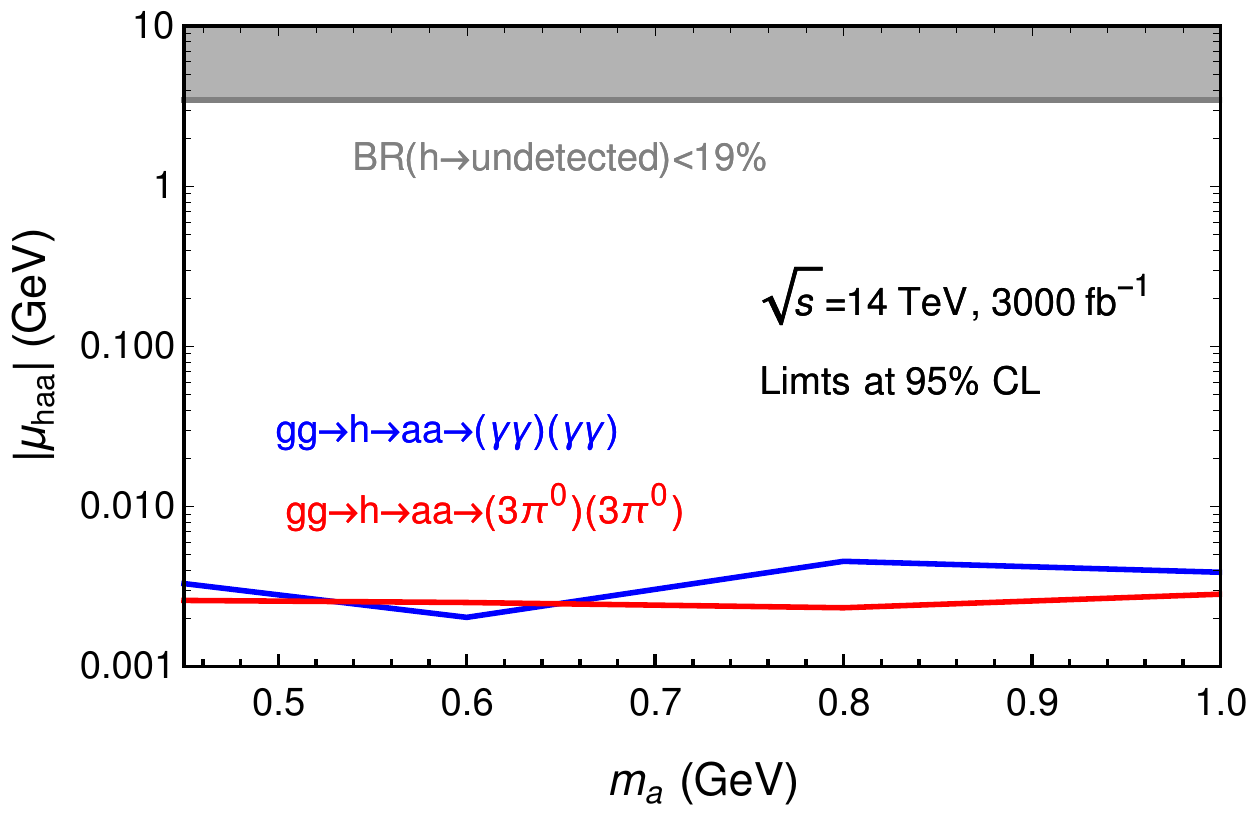}}
\end{center}
\caption{\label{fig:sensitivity} The projection of limits at 95\% CL for (a)  $|\mu_{hss}|$ and (b) $|\mu_{haa}|$ for various photon-jet signatures at the LHC with $\sqrt{s}=14$ TeV and an integrated luminosity of $3000\,\textrm{fb}^{-1}$ without the pileup effect.}
\end{figure}

The $h\rightarrow ss\,(aa)$ coupling is parameterized as a dimension-1 coefficient $\mu_{hss}$ ($\mu_{haa}$) and the sensitivities of photon-jets signatures is translated to this parameter. The branching ratio for each channel, i.e.~$s/a\rightarrow\gamma\gamma$, $s\rightarrow\pi^0\pi^0$ and $a\rightarrow\pi^0\pi^0\pi^0$ is assumed to be 1 in a model-independent way. The FeynRules~\cite{Alloul:2013bka} is used to generate the UFO~\cite{DEGRANDE20121201} model files and MadGraph 5~\cite{Alwall:2014hca} is used to generate the signal and background events produced in $pp$ collisions at $\sqrt{s}=14$ TeV. Additional interactions from the pileup effect are not considered in this study. All events are passed to PYTHIA 8~\cite{Sjostrand:2007gs} for parton showering and hadronization.

The generated truth events are required to have at least two photon-jet candidates with $\Delta R_J < 0.25$, $\log\theta_J < -0.8$ and $N_{track}=0$~\cite{Ellis:2012zp,Sheff:2020jyw}, where $\Delta R_J$ is the cone of radius, $\theta_J$ is the hadronic energy fraction of a jet and $N_{track}$ is the number of charged particles inside a jet. The photon-jets are required to be isolated from nearby charged tracks within $\Delta R < 0.2$.
The leading ($J_1$) and sub-leading ($J_2$) photon-jets are required to be boosted with transverse momentum $P_T (J_{1,2}) > 40$ GeV and $|\eta|<2.5$. Moreover, the cuts of $P_T (J_1) > 0.4 M_{J_1 J_2}$, $P_T (J_2) > 0.3 M_{J_1 J_2}$ and $120 < M_{J_1 J_2} < 130$ GeV, where $M_{J_1 J_2}$ is the invariant mass of $J_1$ and $J_2$, are applied to further suppress the SM backgrounds. Finally, the identification efficiencies for the highly boosted signals and SM backgrounds using CNN are applied to obtain the events at reconstruction level assuming the efficiencies are $\eta$-independent. 

Projection of 95\% Confidence Level (CL) limit for $|\mu_{hss}|$ and $|\mu_{haa}|$ at the LHC with $\sqrt{s}=14$ TeV and an integrated luminosity of $3000\,\textrm{fb}^{-1}$ without pileup effect is shown in Fig.~\ref{fig:sensitivity}. The recent Higgs boson exotic decays constraint measured in $pp$ collision data at $\sqrt{s}=13$ TeV using an integrated lumininosity of $139\,\textrm{fb}^{-1}$, $B(h\rightarrow \textrm{undetected}) < 19\%$~\cite{ATLAS:2020qdt}, is added for the comparison. With the good performance of CNN, the $2\sigma$ bounds for $|\mu_{hss}|$ and $|\mu_{haa}|$ can reach $\cal O$(1) MeV.

\section{Conclusions}
Deep learning technique has been applied to identify the photon-jet signatures produced through the Higgs boson exotic decays. The results show that the CNN is a promising tool to separate the photon-jet signatures from SM backgrounds such as the single photon and $\pi^0$ from QCD jets. Using CNN, the photon-jet can be identified with efficiency above 99.2\% with the backgrounds rejection rate above 99.8\%. For photon-jet with energy above 150 GeV, CNN shows profound improvement with respect to BDT based on shower shape variables.

The sensitivity of exploring such Higgs boson exotic decays with novel photon-jet signatures at the LHC without the pileup effect has been studied. The limits on the $h\rightarrow ss\,(aa)$ coupling at 95\% CL as a function of the mass of $s\,(a)$ have been obtained at the 14 TeV LHC with an integrated luminosity of $3000\,\textrm{fb}^{-1}$.

\ack
The work of S.-C. Hsu and K. Li are supported by the U.S. Department of Energy, Office
of Science, Office of Early Career Research Program under Award number DE-SC0015971. C.-T. Lu is supported by KIAS Individual Grant No. PG075301. X. Ai acknowledges support from DESY (Hamburg, Germany), a member of the Helmholtz Association HGF.

\section*{References}
\bibliography{reference.bib}

\end{document}